\documentclass[prl,amsmath,amssymb,twocolumn]{revtex4-1} 
\usepackage{graphicx}
\usepackage[table]{xcolor}
\usepackage[english]{babel}

\begin{document}
          
\title{Band structures of plasmonic polarons}

\author{Fabio Caruso}
\author{Henry Lambert}
\author{Feliciano Giustino}
\affiliation{Department of Materials, University of Oxford, Parks Road, Oxford OX1 3PH, United Kingdom}
\date{\today}

\date{\today}

\begin{abstract}
Using state-of-the-art many-body calculations based on the 
`\textit{GW} plus cumulant' approach, we show that 
electron-plasmon interactions lead to the emergence of plasmonic 
polaron bands in the band structures of common semiconductors. 
Using silicon and group IV transition 
metal dichalcogenide monolayers (AX$_2$ with A=Mo,W and X=S,Se)
as prototypical examples, we demonstrate that these
new bands are a general feature of systems
characterized by well defined plasmon resonances. 
We find that the energy vs.\ momentum dispersion relations 
of these plasmonic structures closely follow the 
standard valence bands, although they appear 
broadened and blueshifted by the plasmon energy. 
Based on our results we identify general 
criteria for observing plasmonic polaron bands in 
the angle-resolved photoelectron spectra of solids. 
\end{abstract}

\maketitle   
{Electron-boson interactions are pervasive in many-body physics} 
and the resulting quasiparticles are clear examples of emergent behavior in
quantum matter.  
While in the case of phonons and magnons the fingerprints of their interactions
with electrons in angle-resolved photoemission spectra (ARPES) are largely understood~\cite{Damascelli2003},
much less
is known about electron-plasmon interactions. 
{A detailed and quantitative description of these interactions 
is key to refine our understanding of electronic excitations in 
condensed matter, and 
could provide new pathways towards plasmon-assisted band-gap tuning \cite{Liang2015}, or 
the manipulation of plasmon polaritons, with potential implications for photonics and 
plasmonics~\cite{Meier2005}. }

In ARPES, 
the acceleration of a photo-electron upon photon absorption
may trigger shake-up excitations in the sample, leading to the emission of phonons,
electron-hole pairs, and plasmons, the latter being collective charge-density fluctuations
\cite{1402-4896-21-3-4-039}. 
Intuitively, if a photon excites both a hole and a plasmon,
the ARPES signal should exhibit spectral weight at energies
corresponding to the sum of the binding energy of the electron and the excitation energy of the plasmon,
as obtained for example from electron energy loss spectroscopy (EELS) \cite{PhysRev.113.1254}.
This phenomenology is analogous to the emergence of `peak-dip-hump' structures in ARPES 
as a result of electron-phonon interactions 
\cite{Lanzara2001, Bostowick2007, Giustino2008, Damascelli2003, Dahm2009, Moser2013, Carbotte2011};
the difference between the resulting spectral features arises from
the characteristic energy of the boson ($\sim$10~meV for phonons, $\sim$10~eV for plasmons).
In fact the very first model of electron-plasmon interactions \cite{Lundqvist, PhysRevB.1.471,
1402-4896-21-3-4-039} is formally equivalent to the electron-phonon Hamiltonian developed for the polaron
problem \cite{Engelsberg1963}. In this model, the electron-plasmon interaction
results in `plasmonic polarons' \footnote{We purposely avoid the term ``plasmarons'' introduced
by Lundqvist \cite{Lundqvist}. In the original formulation the plasmaron was understood
as a new elementary excitation resulting from a pole of the electron's Green's function. Subsequent
analysis by Langreth \cite{PhysRevB.1.471} showed that the poles identified in
\cite{Lundqvist} are an artifact of the $GW$ approximation, but there exists a polaron
resonance in analogy with the standard electron-phonon theory \cite{PhysRevB.1.471,
1402-4896-21-3-4-039}.}, in complete analogy with the polarons of the ordinary theory of electron-phonon
interactions \cite{mahan2000many}.

Identifying plasmonic polarons in ARPES spectra is notoriously difficult. 
While
plasmonic satellites have been successfully identified in the {\it integrated} photoemission
spectra of Na~\cite{PhysRevLett.77.2268} and Si~\cite{PhysRevLett.107.166401},
the identification of energy  vs.\ momentum dispersions 
of plasmonic polarons
in {\it angle-resolved} spectra has proven considerably more challenging \cite{Kheifets2003}.
So far such dispersions have been observed only in the case of graphene, and only in a narrow region
of the Brillouin zone around the Dirac point~\cite{Bostwick21052010,PhysRevB.77.081411}.
Key factors hindering the observation of the dispersion relations of plasmonic polarons are
(i) the energy scale of the plasmon energy, which requires using energetic photons at the expense
of momentum resolution; (ii) the increased phase-space for
electron-phonon scattering and electron-hole pair generation, which adds to the spectral broadening;
and (iii) the possible mix up of weak plasmonic satellites and strong quasi-particle peaks.

In this work we perform state-of-the-art first-principles calculations
to show that electron-plasmon interactions lead to the formation 
of plasmonic-polaron band-structure replica.
These new structures appear as broadened copies of the valence bands shifted by the plasmon energy.
Using a combination of many-body perturbation theory in the $GW$ 
approximation \cite{PhysRev.139.A796,Hybertsen1986, Onida2002}
and the cumulant expansion approach~\cite{PhysRevB.1.471,1402-4896-21-3-4-039,PhysRevLett.77.2268,Holm1997, PhysRevLett.107.166401, PhysRevLett.110.146801,Guzzo2012,Kas2014,Guzzo2014}
we demonstrate the presence of plasmonic-polaron bands in silicon.
We {further} show that 
two-dimensional group IV transition-metal dichalcogenides 
(TMDs) AX$_2$ with A=Mo,W and X=S,Se \cite{Radisavljevic2011, Wang2012} provide an ideal 
playground for the experimental observation of these 
novel spectroscopic signatures of the electron-plasmon coupling.

Within the sudden approximation, the photocurrent measured in ARPES experiments is proportional 
to the electron spectral function $A({\bf k},\omega)$ \cite{0953-8984-11-42-201,Damascelli2003}, 
where ${\bf k}$ is the crystal momentum of the electron
and $\omega$ its binding energy (here and in the following atomic units 
are understood). The spectral function can be calculated by using the cumulant 
expansion~\cite{PhysRevB.1.471,1402-4896-21-3-4-039,PhysRevLett.77.2268, PhysRevLett.107.166401, 
PhysRevLett.110.146801}: 
the electron Green's function is expanded in terms of the screened Coulomb 
interaction $W$, and a subset of diagrams is evaluated to all orders 
of perturbation \cite{1402-4896-21-3-4-039}. 
This strategy leads to a more accurate treatment of dynamical correlation as compared to 
the standard $GW$ approximation \cite{PhysRev.139.A796}. 
The cumulant expansion draws from the exact solution of the polaron problem \cite{mahan2000many}, and 
was originally applied to study plasmon satellites in core-level spectra \cite{PhysRevB.1.471}.
Importantly, it is also valid in the case of valence electrons, as the effects 
of electron recoil (change of electron momentum) upon plasmon emission tend to cancel 
out \cite{1402-4896-21-3-4-039}. 

In this work we use the formulation of the cumulant expansion given by \cite{1402-4896-21-3-4-039}
and \cite{PhysRevLett.77.2268}, which we will refer to as the $GW$+C$_{\rm AHK}$ approach. 
In this formulation only the first cumulant is retained in order to describe the lineshape
of quasi-particles and one-plasmon excitations. 
The case of $n$-plasmon excitations ($n\ge 2$) is of little
interest here since the corresponding spectral signatures are damped by the Lang-Firsov
factor $a^n/n!$, with $a$ the average number of plasmons around the hole \cite{PhysRevB.1.471}.
The $GW$+C$_{\rm AHK}$ spectral function can 
be expressed as \cite{PhysRevLett.77.2268}:
  \begin{equation}\label{eq-spectrum1}
  A({\bf k},\omega) = \sum_{n} [ A_n^{\rm QP}({\bf k},\omega) + 
  A_n^{\rm QP}({\bf k},\omega)A_n^{\rm C}({\bf k},\omega)]. 
  \end{equation}
Here $A_n^{\rm QP}$ denotes the quasi-particle contribution to the $G_0W_0$ 
spectral function: 
  \begin{equation}\label{eq.specfun}
  A_n^{\rm QP}({\bf k},\omega)  
  = \frac{1}{\pi} \frac{|\Sigma^{\prime\prime}_{n{\bf k}}(\varepsilon_{n{\bf k}})|}
  {[\omega - \varepsilon_{n{\bf k}} - \Sigma^\prime_{n{\bf k}}(\varepsilon_{n{\bf k}})]^2 
  + [\Sigma^{\prime\prime}_{n{\bf k}}(\varepsilon_{n{\bf k}})]^2},
  \end{equation}
where $\Sigma^\prime$ $(\Sigma^{\prime\prime})$ indicates the real (imaginary) part 
of the $G_0W_0$ self-energy \cite{PhysRev.139.A796,Hybertsen1986}, 
and $\varepsilon_{n{\bf k}}$ the Kohn-Sham eigenvalue.
In Eq.~(\ref{eq.specfun}) it is assumed 
that the off-diagonal elements of the self-energy are small and can be neglected, as is typically 
the case \cite{Hybertsen1986, Bruneval2006}. 
The term $A_n^{\rm C}$ in Eq.~(\ref{eq-spectrum1}) 
is defined as \cite{PhysRevLett.77.2268}: 
  \begin{equation}\label{eq-spectrum2}
  A_n^{\rm C}({\bf k},\omega)  
  =  \frac{\beta_{n{\bf k}}(\omega) - \!\beta_{n{\bf k}}(\varepsilon_{n{\bf k}}) - 
  \!(\omega-\varepsilon_{n{\bf k}})\!\left. \displaystyle\frac{\partial \beta_{n{\bf k}}}{\partial \omega}
    \right|_{\varepsilon_{n{\bf k}}}}
  {(\omega-\varepsilon_{n{\bf k}})^2},
  \end{equation}
where $\beta_{n{\bf k}}(\omega) = \pi^{-1}\Sigma_{n{\bf k}}^{\prime\prime}(\omega)
\theta(\mu-\omega)$, 
$\mu$ being the chemical potential.
This term accounts for interactions between the photo-hole and one-plasmon 
excitations \cite{0953-8984-11-42-201}, and its contribution to the
spectral function in Eq.~(\ref{eq.specfun}) is to be identified with plasmonic polarons.

  \begin{figure}[t]
  \begin{center}
  \includegraphics[width=0.48\textwidth]{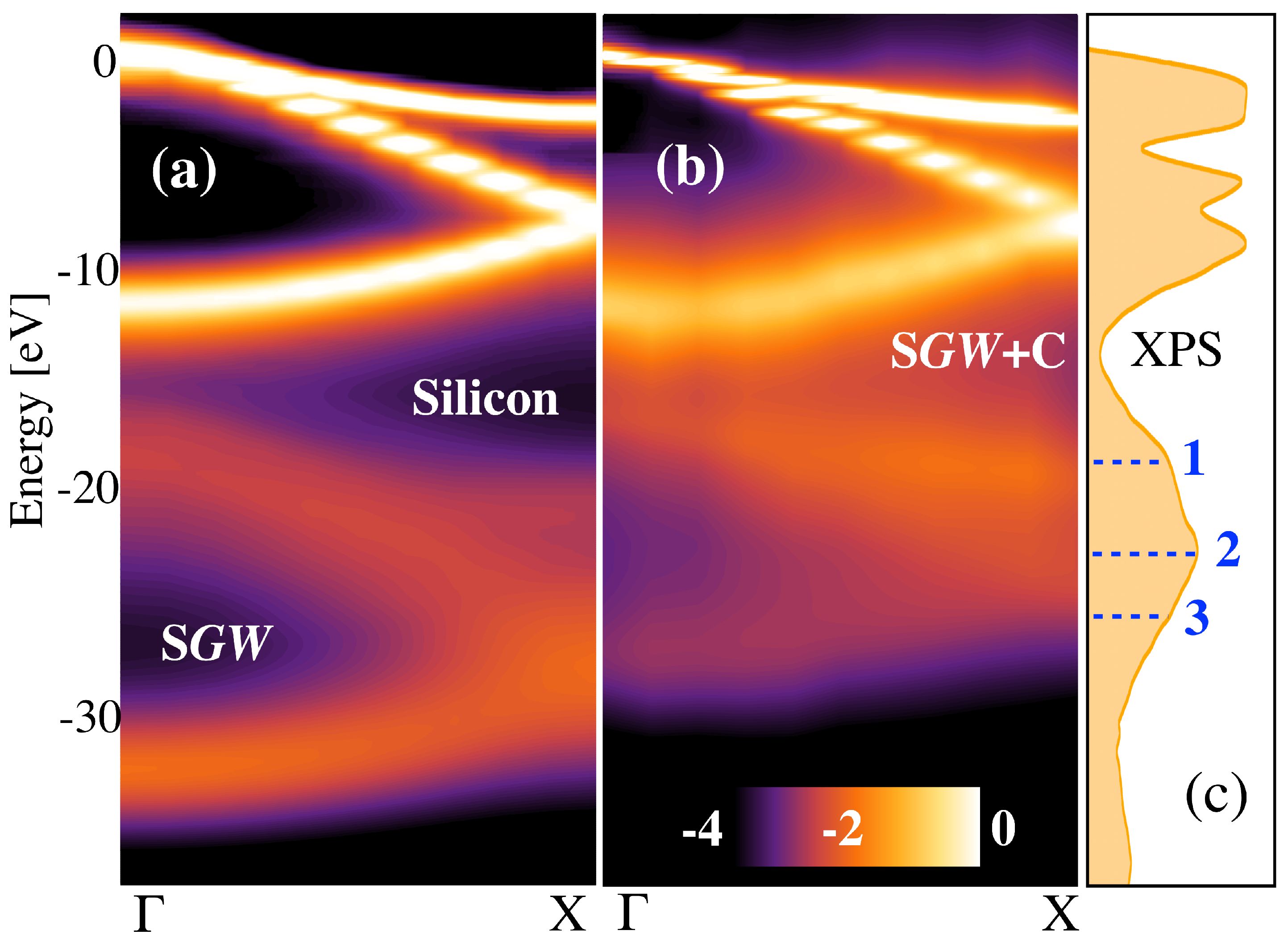}
  \end{center}
  \caption{\label{fig-si}
  Angle-resolved spectral function of silicon on a logarithmic scale
  for wavevectors along the $\Gamma$-$X$ high-symmetry line,
  evaluated using (a) the Sternheimer-$GW$ method (S$GW$) and (b) the S$GW$ plus cumulant (S$GW$+C) 
  approach.
  (c) Measured X-ray photoemission spectrum of silicon (XPS) from
  Ref.~\cite{PhysRevLett.107.166401}. The blue dashed lines indicate the three features 
  discussed in the main text.}
  \end{figure}

\begin{figure*}
  \begin{center}
  \includegraphics[width=0.98\textwidth]{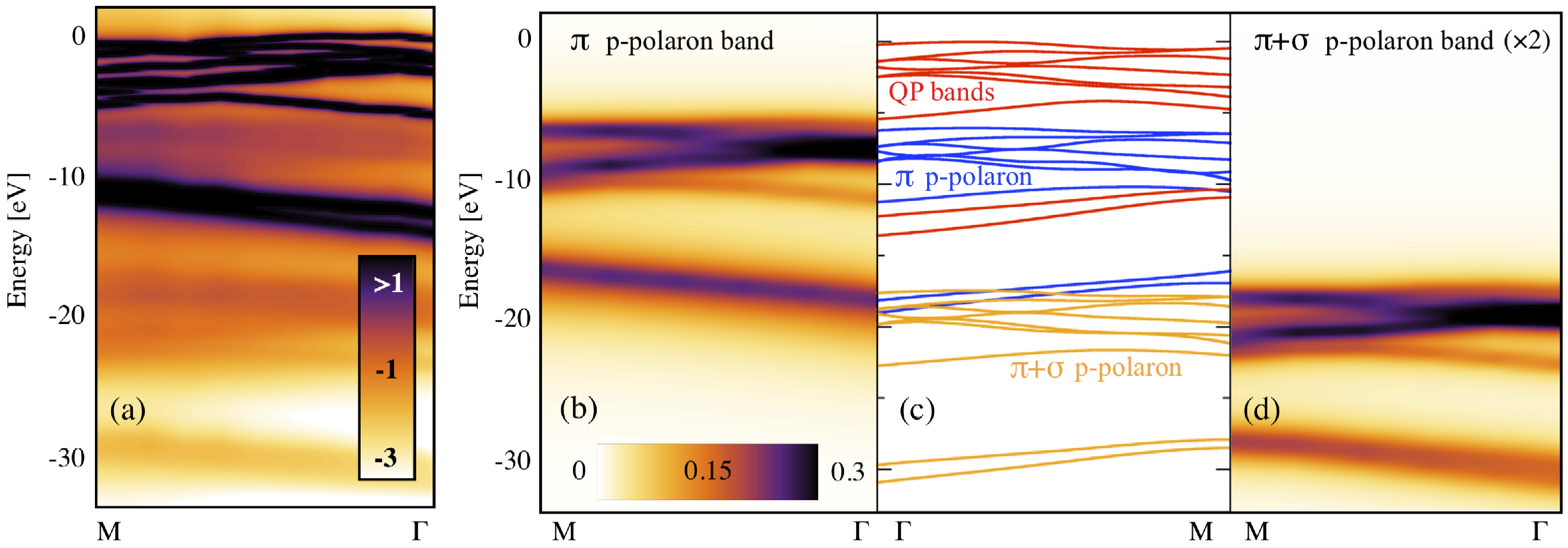}
  \end{center}  
  \caption{\label{fig-plasmaron-bands}
  (a) Complete angle-resolved spectral function of {monolayer} MoS$_2$ on a logarithmic scale
  for wavevectors along the $\Gamma$-$M$ high-symmetry line, 
  evaluated using the $GW$+C$_{\rm AHK}$ approach.
  (b), (d) Contributions of plasmonic polarons to the spectral function in (a):
  $\pi$-plasmonic polarons (b) and $({\pi+\sigma})$-plasmonic polarons (d).
  (c)~Quasi-particle band structure of {monolayer} MoS$_2$ extracted from (a) (red solid lines),
  band structure of $\pi$-plasmonic polarons extracted from (b) (blue solid line), and
  band structure of $({\pi+\sigma})$-plasmonic polarons from (d) (yellow solid line).
  Energies are referenced to the valence-band top.
  }
  \end{figure*}

Using Eqs.~(\ref{eq-spectrum1})-(\ref{eq-spectrum2}), we now investigate the signatures
of plasmonic polarons in silicon. 
In Fig.~\ref{fig-si}(a) and (b) we report the angle-resolved spectral function 
of silicon obtained from the Sternheimer-$G_0W_0$ approach (S$GW$) and S$GW$
plus cumulant (S$GW$+C$_{\rm AHK}$), respectively. 
Details on S$GW$ are provided in the Supplemental Material \cite{sup} 
and Refs.~\cite{PhysRevB.81.115105,PhysRevB.88.075117}.
For silicon, the experimental (integrated) photoemission spectrum shown in Fig.~\ref{fig-si}(c)
is characterized by a broadened plasmonic resonance 
covering approximately the energy range from 16~eV to 30~eV below the Fermi energy. 
In this resonance we can clearly identify three distinct structures (dashed blue lines).
The bright energy bands visible 
in Fig.~\ref{fig-si}(a) for binding energies between 0-12~eV correspond 
to the standard quasi-particle peaks. These peaks result from photoionization processes occurring 
in absence of plasmon excitations, and define the ordinary valence band structure of silicon.
In addition to the quasi-particle features, the spectral function exhibits a rich structure
at binding energies between 15-30~eV. These structures can be identified
with plasmonic polarons. 
These features are present already at the $G_0W_0$ level,  
however, their energy range is largely overestimated and, thus, 
not compatible with the plasmonic features observed in XPS. 
The inclusion of the cumulant correction in S$GW$+C$_{\rm AHK}$ moves the plasmonic
polaron resonances to lower binding energy, improving the agreement with 
the experimental spectrum significantly.
{This improvement reflects the inclusion of higher-order exchange-correlation diagrams 
in the $GW$+C$_{\rm AHK}$ Green function, as discussed in more detail in the Supplemental Material \cite{sup}.}

Unexpectedly, plasmonic polarons exhibit dispersion relations which follow closely 
the ordinary band structure resulting from the quasi-particle peaks. 
The striking similarity between the dispersion of the valence bands 
and the plasmonic structures suggests that we are looking 
at {\it band structures} of plasmonic polarons. 
Plasmonic polaron bands appear as blue-shifted 
replicas of the ordinary valence bands, but they are 
considerably broader and less intense. 
The comparison of Fig.~\ref{fig-si}(b) and (c) suggests that the plasmon satellite 
of silicon \cite{PhysRevLett.107.166401} results from the momentum-average 
of plasmonic polaron bands over the first Brillouin zone.
{For quasiparticles, it is well known that the density of states is characterized by singularities 
(known as Van Hove singularities) at the energies for which  the first momentum derivative of the quasiparticle 
bands vanishes ($\nabla_{\bf k} \varepsilon_{\bf k} = 0$). 
Correspondingly, peaks in the density of states can be associated with  
extremal points of quasiparticle bands.
Figures~\ref{fig-si}(b) and (c) indicate that Van Hove singularities may also have a plasmonic origin. 
In particular, the two experimental peaks at 20.4~eV and 24.5~eV
and the shoulder at 27.3~eV [labelled as 1--3 in Fig.~\ref{fig-si}(c)]  
can be attributed to the vanishing of the first momentum derivative of 
the plasmonic polaron bands in Fig.~\ref{fig-si}(b).
}

Owing to the large plasmon energy, the experimental observation of
plasmonic polaron bands in silicon may be hindered by the low resolution of ARPES 
measurements well below the Fermi energy. 
In order to identify materials in which such polaron bands may be observed,
in the following we focus on
group IV transition metal dichalcogenides
(MoS$_2$, WS$_2$, MoSe$_2$, and WSe$_2$).
The EELS spectra of the three-dimensional parent 
compounds exhibit two distinct features around $\sim$8~eV and $\sim$22~eV, 
corresponding 
to the excitation of $\pi$ and ${\pi+\sigma}$ plasmons, respectively \cite{doi:10.1080/14786436908225867,Bell1976}. 
Since the width of the bands arising from transition metal $d$ states 
and chalcogen $p$ states in these compounds is approximately 7~eV \cite{Raybaud1997},
possible plasmonic polarons are expected to appear between $\sim$8-15~eV. In this energy 
range the deep S-$3s$ or Se-$4s$ bands, located between 12-15~eV \cite{Raybaud1997},
dominate the spectral function, thereby hindering the identification of plasmonic polarons in ARPES
also in this case.

At variance with this scenario, in the case of monolayer TMDs both 
experimental~\cite{Coleman04022011} and theoretical~\cite{doi:10.1021/nn201698t} studies 
reported plasmonic peaks in the EELS spectra which are strongly red-shifted with respect 
to their bulk counterpart.
For example, in the case of MoS$_2$ and WS$_2$ monolayers the 
$\pi$ plasmons are found at energies around 6 eV. 
We thus expect to observe plasmonic polarons at binding energies between 6-13~eV. 
Since this energy window matches the band structure {\it gap} between the metal-$d$/chalcogen-$p$ 
bands and the chalcogen $s$ bands, such plamonic polarons should be distinctly observable.

To examine this possibility on quantitative grounds, we calculate the $GW$+C$_{\rm AHK}$
spectral functions of TMDs from first principles\footnote{
  Ground-state density-functional theory calculations in the local density approximation (LDA) 
  \cite{PhysRev.140.A1133,PhysRev.136.B864} 
  are performed using the Quantum-ESPRESSO software package~\cite{0953-8984-21-39-395502}. 
  We used  normconserving Hartwigsen-Goedeker-Hutter pseudopotentials and a 29~Ry  
  kinetic energy cutoff, which suffices for converging the
  quasiparticle energies of MoS$_2$ \cite{PhysRevB.88.245309}.  
  The Brillouin zone is sampled using a 20$\times$20 Monkhorst-Pack grid. 
  For monolayer MoS$_2$ we employed a supercell approach with a 16.4 \AA\ interlayer 
  spacing between periodic replicas. 
  In the $GW$+C$_{\rm AHK}$ spectral function [Eqs.~(\ref{eq-spectrum1}-\ref{eq-spectrum2})] 
  we neglect the asymmetry factor \cite{1402-4896-21-3-4-039} 
  and the exchange part of the quasi-particle correction
  as we are primarily interested in the intensity and binding energy 
  of the spectral features.
  Sternheimer-$GW$ calculations for silicon employ the same computational 
  parameters as in \cite{PhysRevB.88.075117}.}.
In order to contain the computational cost we describe the screening by introducing a two-pole 
approximation for the inverse dielectric matrix, as discussed in the Supplemental
Material \cite{sup}.

Figure~\ref{fig-plasmaron-bands}(a) shows the complete $GW$+C$_{\rm AHK}$ spectral function
of {monolayer} MoS$_2$ evaluated along the $\Gamma$-$M$ high-symmetry line. 
At binding energies between 0-5~eV and 10-15~eV, 
the spectral function of {monolayer} MoS$_2$  exhibits
the standard quasi-particle peaks.
As in the case of silicon, in monolayer MoS$_2$ plasmonic polarons 
introduce {\it new} spectral features in a binding energy range 
where quasi-particle states are absent. 
To characterize these new features, we analyze their energy vs.\ wavevector dispersions
using a Lorentzian decomposition of the energy profiles. 
This analysis allows us to disentangle the contributions 
of plasmonic polarons from the quasi-particle excitations. 
Figures~\ref{fig-plasmaron-bands}(b) and (d) 
show the plasmonic polarons corresponding to the emission of a photo-electron and the simultaneous 
excitation of a $\pi$ or ${\pi+\sigma}$ plasmon, respectively. 
As for silicon, plasmonic polarons exhibit 
a clear energy-momentum dispersion 
relation that leads to the formation of blue-shifted replica of 
the valence band structure.
In particular we find two replicas of the valence bands of 
{monolayer} MoS$_2$, one associated with the $\pi$ plasmon centered around 
the binding energy $\sim$8~eV, and another one associated with the ${\pi+\sigma}$ plasmon 
around $\sim$19~eV. 

Owing to the approximate treatment of electron recoil effects 
in the cumulant approach \cite{1402-4896-21-3-4-039,PhysRevLett.107.166401}, 
the actual broadening might be even larger than in the present calculations.
In order to understand whether plasmonic polaron bands 
could be observed in ARPES experiments it 
is therefore essential to quantify their spectral weight. 
A reliable measure of the intensity of the plasmonic bands can be obtained from the average number 
$a_{n{\bf k}}$ of plasmons emitted during the photoemission process~\cite{0953-8984-11-42-201}:
$a_{n{\bf k}} = \int \omega^{-2}\beta_{n{\bf k}}(\omega) d\omega.$
In the case of the high-lying Mo-4$d$/S-3$p$ bands we obtain $a_{n{\bf k}}$ in the range
0.08-0.12 for momenta along the $\Gamma$-$M$ line, whereas $a_{n{\bf k}}$ is found in the range
0.04-0.06 for the lower-lying S-3$s$ bands. 
These results indicate that the formation 
of plasmonic polarons provides an important dissipation channel for the ARPES photo-current.
In particular, in the case of {Mo-4$d_{z^2}$} states at the top of the valence band
(the most important for electron transport in p-doped MoS$_2$) every photo-hole
is accompanied by 0.13 plasmons per unit cell.
These estimates are confirmed by a direct integration 
of the spectral function, which shows that plasmonic 
polarons carry on average 12\% of the total
spectral weight of the valence electronic states.
For the valence states the
intensity ratio between the plasmonic polaron band and the quasi-particle peak 
is $\sim0.01$-$0.02$.
In addition, by considering the metal-$d$/chalcogen-$p$ bands altogether we find that, 
in the case of monolayer MoS$_2$, it should
be possible to extract up to $1.2\cdot10^{14}$ electrons/cm$^2$ (0.1 electrons/cell)
at binding energies inside the band structure gap between Mo-4$d$/S-3$p$ bands and S-3$s$ bands.

\begin{figure}[t]
  \begin{center}
  \includegraphics[width=0.45\textwidth]{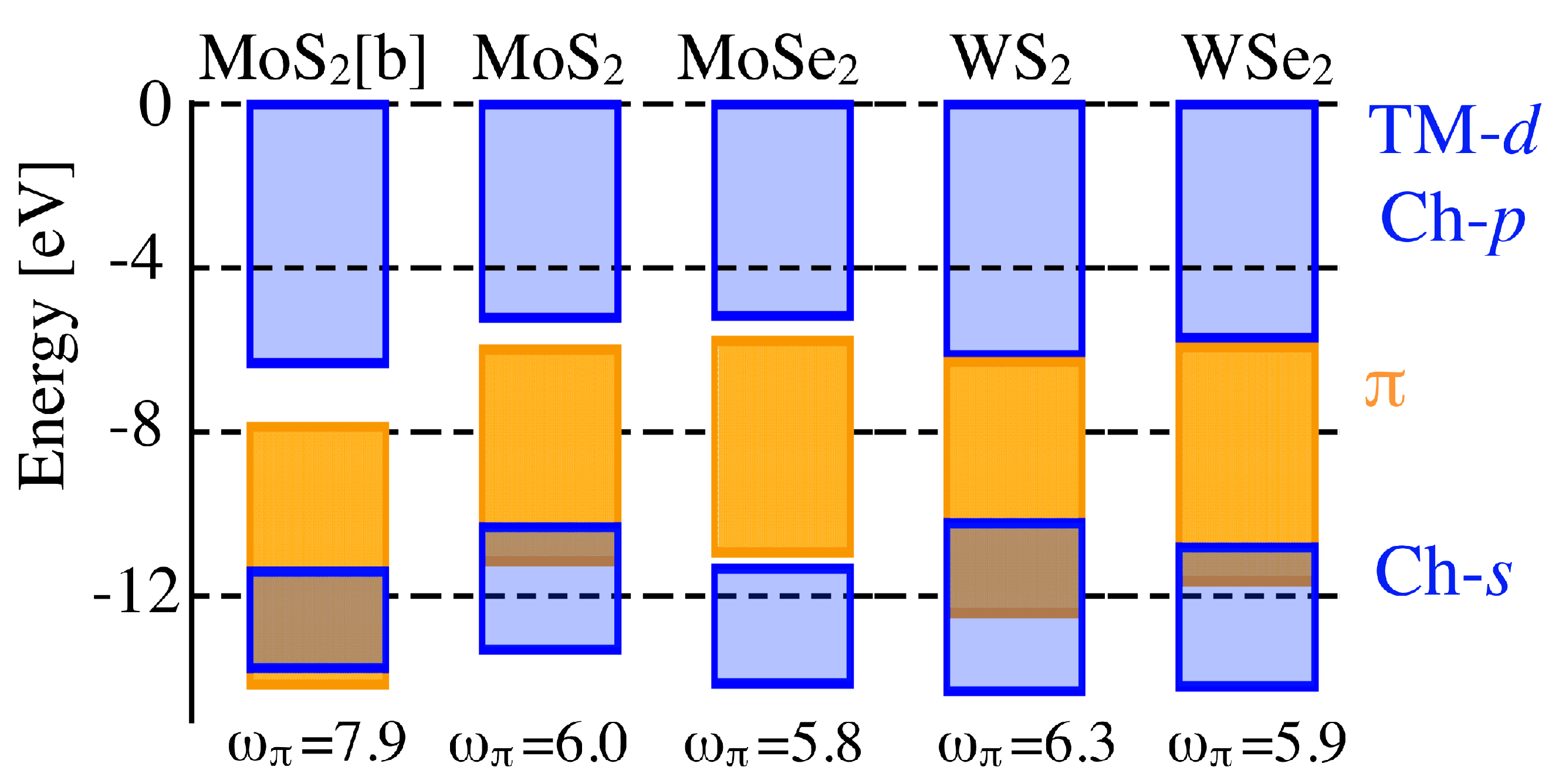}
  \end{center}
  \caption{\label{fig-plasmaron-range}
  Energy range of the quasi-particle bands 
  (blue boxes) and the $\pi$-plasmonic polaron bands in
  {monolayer} MoS$_2$, MoSe$_2$, WS$_2$, and WSe$_2$, evaluated within the $GW$+C$_{\rm AHK}$ approach
  (yellow boxes). Bulk MoS$_2$ is included for comparison (MoS$_2$[b]).
  The topmost bands arise from the hybridization of transition metal $d$ states
  and the chalcogenide $p$ states (TM-$d$/Ch-$p$);
  the low-energy ones from the chalcogenide $s$ states (Ch-$s$).
  The plasma energies $\omega_{\pi}$ are given in eV. 
  }
  \end{figure}

At variance with the case of monolayer MoS$_2$, for the bulk compound 
our calculations indicate that the plasmonic bands overlap 
substantially with the deep S-$s$ states (Fig.~\ref{fig-plasmaron-range}). 
The overlap between ordinary quasi-particle bands and plasmonic polaron 
bands should make the experimental detection of these new features in bulk MoS$_2$ much more challenging
than in the case of a monolayer.

The above results suggest that two key conditions need to be satisfied
for plasmonic polaron band structures to be clearly observed in ARPES: (i) the existence 
of low-energy plasmon excitations in the EELS spectrum, and (ii) the presence of a band gap 
in the valence band manifold. 
The condition (ii) also contributes to minimize 
spectral broadening arising from electron-phonon scattering.

Given these `design rules' it is natural to ask whether there exist `optimal' TMDs
for observing plasmonic polaron bands.
To answer this question, we repeated
our $GW$+C$_{\rm AHK}$ calculations for the related {monolayer} compounds MoSe$_2$, WS$_2$, and WSe$_2$.
Figure~\ref{fig-plasmaron-range} shows that plasmonic polaron bands associated with $\pi$
plasmons fit inside the band gap betwen the metal-$d$/chalcogen-$p$ bands and the chalcogen-$s$ bands
for all these compounds. In particular we find that {monolayer} MoSe$_2$ should provide an ideal
testbench for identifying plasmonic polarons, since in this case the plasmonic structures
exhibit essentially no overlap with the ordinary valence bands (Fig.~\ref{fig-plasmaron-range}).

Since the plasmon energy scales with square root of the static dielectric constant $\epsilon_0$,
it should also be possible to realize plasmonic band structure engineering in TMD monolayers 
by modifying their dielectric screening properties. For example this could be achieved by using 
different substrates, by building van der Waals heterostructures of TMDs \cite{Geim2013},
using doping \cite{Radisavljevic}, 
or mechanical deformation \cite{doi:10.1021/nl402875m,
PhysRevB.86.241401}. All these modifications would leave the ordinary valence band structure 
essentially unaffected (on the eV scale), while producing significant shifts in the binding energy 
of plasmonic polarons.

In conclusion, using state-of-the-art first-principles $GW$ plus cumulant calculations, we have shown
that electron-plasmon coupling leads to the formation of plasmonic polaron band structures,
and that group IV transition metal dichalcogenide monolayers (in particular {monolayer} MoSe$_2$)
should provide a unique opportunity for observing these new features in ARPES experiments. 
Remarkably, we have found that these plasmonic band structures 
exhibit dispersion relations which closely follow the ordinary valence bands {and,
similarly to quasiparticle bands, lead to the formation of Van Hove singularities in the 
density of states.}
{More generally, plasmonic-polaron bands emerge as novel spectroscopic 
signatures of electron-plasmon coupling, which may contribute 
to unravel the complexity of ARPES measurements.}
The emergence of plasmonic polarons is not limited to semiconductors, 
and may also prove useful to rationalize the electronic 
structure of materials characterized by well defined 
plasmonic excitations, including for example metals, as well as $d$- and $f$-electron 
systems \cite{Gatti2013, Basak2009}.
In fact, our work raises the question on whether 
the concept of plasmonic polarons may be systematically employed 
in the interpretation of ARPES spectra of complex systems. 
As an example, while our work focused on the simplest case of `isolated' plasmonic polaron bands, 
the crossing of plasmonic bands and high-lying valence bands
may reveal the emergence of band-branching effects,
in analogy with the polaron problem in electron-phonon physics \cite{Eiguren2009}.

\acknowledgments
This work was supported by the Leverhulme Trust (Grant RL-2012-001) and the European Research Council
(EU FP7 / ERC grant no. 239578 and EU FP7/grant no. 604391 Graphene Flagship). Calculations were
performed at the Oxford Supercomputing Centre and at the Oxford Materials Modelling Laboratory.


%

\end{document}